\documentclass[11pt,preprintnumbers,aps,amssymb,nofootinbib,amsmath]
{revtex4}
\usepackage{epsfig,epsf}
\usepackage{xfrac} % makes nice fractions with \sfrac{ }{ } command (in or out of Math mode)
\usepackage{bm} % puts greek and math symbols in boldface using \bm
\newcommand{\beq}{\begin{equation}}
\newcommand{\beql}[1]{\begin{equation}\label{#1}}
\newcommand{\eeq}{\end{equation}}
\newcommand{\eq}[1]{(\ref{#1})}
\newcommand{\fig}[1]{Fig.~\ref{#1}}
\renewcommand{\sec}[1]{Sec.~\ref{#1}}
\newcounter{topiccounter}
\renewcommand{\b}[1]{{\bm #1}} 
\newcommand{\unit}[1]{\hat {{\bm #1}}} % unit vector

\newcommand{\as}{\alpha_s}

\newcommand{\e}{\varepsilon}
\newcommand{\aver}[1]{\left\langle #1 \right\rangle}
\newcommand{\jpsi}{J\mskip -2mu/\mskip -0.5mu\psi}

\begin{document}

\title{$\jpsi$ dissociation in  parity-odd  bubbles}

\author{Kirill Tuchin\\}

\affiliation{
Department of Physics and Astronomy, Iowa State University, Ames, IA 50011}

\date{\today}


\begin{abstract}

We calculate the quarkonium dissociation rate in the $P$ and $CP$-odd domains (bubbles) that were possibly  created in heavy-ion collisions. In the presence of the magnetic field produced by the valence  quarks of colliding ions, parity-odd domains generate electric field.  Quarkonium dissociation is the result  of quantum tunneling of quark or antiquark through the potential barrier in this electric field. The strength of the electric field in the quarkonium comoving frame depends on the quarkonium velocity with respect to the background magnetic field.  We investigate momentum, electric field strength and azimuthal dependence of the dissociation rate.  Azimuthal distribution of quarkonia surviving in the electromagnetic field is strongly anisotropic; the form of anisotropy  depends on the relation between the electric and magnetic fields and quarkonium momentum $P_\bot$. These features can be used to explore the properties of the electromagnetic field created in heavy ion collisions.

\end{abstract}

\maketitle

%%%%%%%%%%
\section{Introduction}\label{sec:intr}

Solid theoretical arguments \cite{Kharzeev:2007jp} and numerical calculations \cite{Skokov:2009qp,Voronyuk:2011jd} indicate a possible existence of very strong magnetic fields in heavy-ion collisions.  Electromagnetic fields of such enormous intensity have never been experimentally studied  and therefore  present a great interest, which  extends far beyond applications in the nuclear physics. What are the possible manifestations of such magnetic field? An effect that has recently attracted a lot of attention is the Chiral Magnetic Effect (CME) \cite{Kharzeev:2004ey,Kharzeev:2007jp,Kharzeev:2007tn,Fukushima:2008xe,Kharzeev:2009fn}.
If a metastable $P$ and $CP$-odd bubble is induced by axial anomaly in the hot nuclear matter, then in the presence of external magnetic field $\bm B_0$ the bubble generates an electric field which is \emph{parallel} to the magnetic one. According to \cite{Kharzeev:2007tn} the value of the electric field $\bm E_0$ in the bubble is 
\begin{align}\label{cme}
\bm E_0 = - N_c\sum_f\frac{e_f^2}{4\pi^2}\, \frac{\Theta}{N_f}\, \bm B_0 = -\frac{2}{3}\frac{\alpha\, \Theta}{\pi}\bm B_0
\end{align}
where the sum runs over quark flavors $f$ and we assumed that only three lightest flavors contribute. The value of the $\Theta$-angle fluctuates from event to event.  CME refers to the  macroscopic manifestation of this effect --  separation of electric charges with respect to the reaction plane. This effect is a possible explanation of  experimentally observed charge asymmetry fluctuations \cite{:2009uh,:2009txa,Ajitanand:2010rc}.

Other effects of the magnetic field that \emph{do not require existence of the  parity-odd bubbles} and have been recently discussed  are:  synchrotron radiation by fast fermions \cite{Tuchin:2010vs}, polarization of the fermion spectra  \cite{Tuchin:2010vs} (also known as the Sokolov-Ternov effect \cite{Sokolov:1963zn}),  enhanced dilepton production \cite{Tuchin:2010gx} and  azimuthal anisotropy of the quark-gluon plasma (QGP) \cite{Mohapatra:2011ku,Tuchin:2011jw}.  These possible effects await their experimental study. Thus, strong magnetic field must have a powerful impact on the behavior of the quark-gluon plasma.

\bigskip

Since CME cannot be the only effect induced by the magnetic field we are motivated to look for magnetic field manifestations in other observables.
We have recently pointed out in Ref.~\cite{Marasinghe:2011bt} that the Lorentz ionization, i.e.\ dissociation of a moving bound state in external magnetic field, is phenomenologically significant in relativistic heavy-ion collisions in the presence of  strong \cite{Kharzeev:2007jp,Skokov:2009qp} quasi-static \cite{Tuchin:2010vs} magnetic field  generated by colliding ions. The ionization or, equivalently, dissociation happens because  quarkonium constituents have finite probability to tunnel through the potential barrier 
in the presence of  electric field, which appears upon boosting to the comoving frame. 
The dissociation rate depends on the magnetic field strength, bound state velocity and its binding energy. The dissociation rate indirectly depends on the  properties of the nuclear matter by the way of dependence of the binding energy on temperature. Since at higher temperatures the binding energy is smaller, the dissociation rate is higher. Still, the Lorentz ionization can happen even if no matter is formed, provided that the magnetic field is strong enough and/or the bound state is fast enough. 

In this paper we address a different type of ionization which is possible only if a metastable $P$ and $CP$-odd bubbles are formed and trigger  emergence of electric field $\bm E_0$ given by \eq{cme}. Due to this field, there is a finite dissociation rate of quarkonium \emph{at rest} in the laboratory frame. When boosted to the comoving frame  quarkonium dissociation is a combined effect 
of the boosted electric  $\bm E_0$ and magnetic $\bm B_0$ fields. In the comoving frame, electric and magnetic fields are directed at some angle with respect to each other depending on the quarkonium kinematics. The main goal of this paper is to investigate the dissociation rate in this case. 

%%%%%%%%%%
\section{Dissociation rate}\label{secII}

Ionization probability of quarkonium equals its tunneling probability through  the potential barrier. In the WKB approximation the later is given by the transmission coefficient and was calculated in \cite{Marasinghe:2011bt}. 
In this method contribution of the quark spin can be easily taken into account. 
Another method of calculating the ionization probability, the imaginary time method  \cite{ITM1,ITM2}, was employed in \cite{Popov:1997-A,Popov:1998-A,Popov:1998aw} by Popov, Karnakov and Mur. In particular, they derived in the non-relativistic approximation the pre-exponential factor that appears due to the deviation of the quark wave function from the quasi-classical approximation. Such a calculation requires matching quark wave function inside and outside the potential barrier \cite{LL3-77}. Extension of this approach to the relativistic case is challenging due to analytical difficulties of the relativistic two-body problem. Fortunately, it was argued in \cite{Popov:1998aw,Marasinghe:2011bt} that the non-relativistic approximation provides a very good accuracy in the $\e_b\ll m$ region which is relevant in the quarkiononium dissociation problem. 

Magnetic field produced in heavy-ion collisions may have a complicated spatial and temporal structure (see e.g.\ \cite{Ou:2011fm}). However, we will assume that the field is constant on the scales relevant for the problem of $\jpsi$ dissociation.  This approximation is supported by our recent  arguments \cite{Tuchin:2010vs} that  the relaxation time of magnetic field is of the order of a few fm/c due to high electrical conductivity of the QGP\footnote{Calculations in Refs.~\cite{Kharzeev:2007jp,Skokov:2009qp,Voronyuk:2011jd} yield very short relaxation time of magnetic field because they neglect the electromagnetic response of the quark-gluon medium, see \cite{Tuchin:2010vs} for details.}. Also, spatial inhomogeneity in the transverse plane reveals itself at distances of the order of the nucleon size and perhaps even larger in view of uniformity of the matter distribution in the nuclei. A more quantitative estimate of the role of spatial and temporal dependence of the magnetic field on $\jpsi$ dissociation requires numerical solution of magneto-hydrodynamic equations and is beyond the scope of this paper.

Given the  electromagnetic field in the laboratory frame $\bm B_0$, $\bm E_0$, the electromagnetic field $\bm B$, $\bm E$ in the comoving frame moving with velocity $\bm V$   is given by 
\begin{subequations}\label{em-cf-0}
\begin{align}
\bm E=& E_0\left\{ \gamma_L(\bm b_0+\rho_0^{-1}\bm V\times \bm b_0)- (\gamma_L-1)\bm V\frac{\bm V\cdot \bm b_0}{V^2}\right\}\\
\bm B=& B_0\left\{ \gamma_L(\bm b_0-\rho_0\bm V\times \bm b_0)- (\gamma_L-1)\bm V\frac{\bm V\cdot \bm b_0}{V^2}\right\}
\end{align}
\end{subequations}
where $\bm b_0=\bm B_0/B_0$ is a unit vector in the magnetic field direction, $\rho_0 = E_0/B_0= 2\alpha |\Theta|/3\pi$ (see \eq{cme}) and $\gamma_L= 1/\sqrt{1-V^2}$. It follows from \eq{em-cf-0} that
\begin{subequations}\label{em-cf}
\begin{align}
E&=E_0\sqrt{1+\gamma_L^2(\bm b_0\times \bm V)^2(1+\rho_0^{-2})}\\
B&= B_0\sqrt{1+\gamma_L^2(\bm b_0\times \bm V)^2(1+\rho_0^2)}
\end{align}
\end{subequations}
Using \eq{em-cf} we find that the angle $\theta$  between the electric and magnetic field in the comoving frame is 
\begin{align}\label{theta}
\cos\theta  = \frac{\bm E\cdot \bm B}{EB} = \frac{1}{\sqrt{[1+\gamma_L^2(\bm b_0\times \bm V)^2(1+\rho_0^{-2}) ][1+\gamma_L^2(\bm b_0\times \bm V)^2(1+\rho_0^2)}]}&
\end{align}
where  we used the relativistic invariance of $\bm E\cdot \bm B$.

It is useful to introduce dimensionless parameters  $\gamma$, $\epsilon$  and $\rho$ as \cite{Popov:1998aw}
\begin{align}\label{param}
\gamma= \frac{1}{\rho}\sqrt{\frac{2\e_b}{m}}\,,\quad \rho= \frac{E}{B}\,,\quad \epsilon = \frac{eE}{m^2}\left( \frac{m}{2\e_b}\right)^{3/2}
\end{align}
where $m$ is quark mass and  $\e_b$ is quarkonium binding energy.
We will treat the quarkonium binding potential in the non-relativistic approximation, which provides a very good accuracy to the dissociation rate  \cite{Popov:1998aw,Marasinghe:2011bt}.
The quarkonium dissociation rate in the comoving frame in the non-relativistic approximation 
 is given by \cite{Popov:1997-A}
\begin{align}\label{w}
w= \frac{8\e_b}{\epsilon}\, P(\gamma,\theta)\,C^2(\gamma,\theta)\, e^{-\frac{2}{3\epsilon}g(\gamma,\theta)} 
\end{align}
where  function $g$ reads
\begin{align}
g&=\frac{3\tau_0}{2\gamma}\left[ 1-\frac{1}{\gamma}\left( \frac{\tau_0^2}{\gamma^2}-1\right)^{1/2}\sin\theta -\frac{\tau_0^2}{3\gamma^2}\cos^2\theta\right] 
\end{align}
and functions $P$ and $C$ are given in the Appendix.  The contribution of quark spin is taken into account by replacing $\e_b\to \e_b'=\e_b-\frac{e}{m}\bm s\cdot \bm B$ \cite{Marasinghe:2011bt}.  Function $g$ represents the leading quasi-classical exponent, $P$ is the pre-factor for the $S$-wave state of quarkonium and $C$ accounts for the   Coulomb interaction between the  valence quarks.  Parameter $\tau_0$ satisfies the following equation 
\begin{align}
\tau_0^2-\sin^2\theta(\tau_0\coth\tau_0-1)^2=\gamma^2
\end{align}
which establishes its dependence on $\theta$ and $\gamma$. Note, that  in the limit $E\to 0$ the dissociation rate \eq{w} exponentially vanishes. This is because  pure magnetic field cannot force a charge to tunnel through a potential barrier.  

Eq.~\eq{w} gives the quarkonium  dissociation rate in a bubble with a given value of $\Theta$. Its derivation assumes that  the dissociation process happens entirely inside a bubble and that $\Theta$ is constant inside the bubble. Since in a relativistic heavy ion collision many bubbles can be produced with a certain distribution of $\Theta$'s (with average $\aver{\Theta}=0$) more than one bubble can affect the dissociation process. This will result in a distractive interference leading to reduction of the $CP$-odd effect on quarkonium dissociation. However, if a typical bubble  size $R_0$ is much larger than the size of quarkonium $R_J$, then  the dissociation is affected by one bubble at a time independently of others, and hence the interference effect can be neglected. In this case \eq{w} provides, upon a proper average, a reasonable estimate of quarkonium dissociation in a heavy ion collision. We can estimate the bubble size as the size of the sphaleron, which is of the order of the chromo-magnetic screening length $\sim 1/g^2T$, whereas the quarkonium size is of the order $\as/\e_b$. Consequently, at small coupling and below the zero-field dissociation temperature  (i.e.\ when $\e_b$ is not too small) $R_0$ is parametrically much larger than $R_J$. A more quantitative estimate of the sphaleron size is   $R_0\simeq 1.2/\as N_c T\simeq 0.4$~fm  \cite{Moore:2010jd}; whereas  for $\jpsi$ $R_J\simeq \as/\e_b\simeq 0.1-0.2$~fm. Thus, based on this estimate bubble interference can be neglected in the first approximation.  However, since the ratio  $R_J/R_0$ is actually not so small this effect nevertheless warrants further investigation. 

To obtain the experimentally observed $\jpsi$ dissociation rate we need to average \eq{w} over the bubbles 
produced in a given event and then over all events. To this end it is important to note that because the dissociation rate depends only on $\rho_0^2$ it is insensitive to the sign of the $\b E_0$ field or, in other words, it depends only on absolute value of $\Theta$ but not on its sign. Therefore, it  stands to reason that although the precise distribution of $\Theta$'s is not known, \eq{w} gives an approximate event average  with parameter $\Theta$ representing a characteristic absolute value of the theta-angle. 

%%%%
\section{Limiting cases}

Before we proceed with the numerical calculations, let us consider for illustration several limiting cases.
 If  quarkonium moves with non-relativistic velocity, then in the comoving frame electric and magnetic fields are approximately parallel $\theta\approx 0$, whereas in the ultra-relativistic case they are orthogonal $\theta\approx \pi/2$, see \eq{theta}. In the later case the electromagnetic field in the comoving frame does not depend on $E_0$ as seen in \eq{em-cf} and therefore the dissociation rate becomes insensitive to the CME. In our estimates we will assume that $\rho_0<1$ which is the relevant phenomenological situation. Indeed, it was proposed in \cite{Kharzeev:2007tn} that  $\rho_0\sim \alpha\ll 1$ produces charge fluctuations with respect to  the reaction plane of the magnitude consistent with experimental data. 

1) $\theta\gtrsim 0$, i.e.\ electric and magnetic fields are approximately parallel. This situation is realized in the following two cases.  (i) Non-relativistic quarkonium velocities: $V\ll \rho_0$ or (ii)  motion of quarkonium at small angle $\phi$ to the direction of the magnetic field $\bm b_0$: $\phi\ll \rho_0/\gamma_LV$. In both cases $E\approx E_0$ and $B\approx B_0$. This is precisely the case where the dissociation rate exhibits its strongest sensitivity to the strength of the electric field $\bm E_0$ generated by the local parity violating QCD effects. Depending on the value of the $\gamma$ parameter  defined in \eq{param} we can distinguish the case of strong electric field $\gamma\gg 1$ and weak electric field $\gamma\ll 1$ \cite{Popov:1998-A}. In the former case, $g= (3/8)\gamma$, $P= (8/e)^{1/2}\gamma e^{-\gamma^2/2}$, $C= e^{\pi \gamma/2}/\gamma$. Substituting into \eq{w} the dissociation rate reads 
\begin{align}\label{gbig}
w= \frac{8\e_b}{\epsilon\gamma}\sqrt{\frac{8}{e}}  e^{-\gamma^2/2}e^{-\frac{\gamma}{4\epsilon}}
= \frac{16\e_b^2 m}{eB_0}\sqrt{\frac{8}{e}}  e^{-\frac{\e_b}{\rho_0^2m}}e^{-\frac{\e_b^2}{\rho_0eE_0}}\,,\quad \gamma\gg 1
\end{align}
In the later case, $g= P=C=1$ and
\begin{align}\label{zz}
w= \frac{8\e_b}{\epsilon}e^{-\frac{2}{3\epsilon}}=\frac{8\e_b m^2}{eE_0}\left(\frac{2\e_b}{m} \right)^{3/2}  
e^{-\frac{2m^2}{3eE_0}\left(\frac{2\e_b}{m} \right)^{3/2}}\,,\quad \gamma\ll 1
\end{align}
where the electromagnetic field in the comoving frame equals one in the laboratory frame as was mentioned before.

2) $\theta\sim \pi/2$, i.e.\ electric and magnetic fields are approximately orthogonal.\footnote{Note, that the limit $\gamma\gg 1$ is different in $\theta = \frac{\pi}{2}$ and $\theta<\frac{\pi}{2}$ cases \cite{Popov:1997-A}. } This occurs for an ultra-relativistic motion of quarkonium $V\to 1$. In this  case 
\begin{align}\label{gg}
B= E= B_0\gamma_L|\bm b_0\times \bm V|\sqrt{1+\rho_0^2}
\end{align}
This case was discussed in detail in our previous paper \cite{Marasinghe:2011bt}. In particular for $\gamma\ll 1$ we get
\begin{align}\label{zz1}
w = \frac{8\e_b m^2}{eE}\left(\frac{2\e_b}{m} \right)^{3/2}  
e^{-\frac{2m^2}{3eE}\left(\frac{2\e_b}{m} \right)^{3/2}}
\end{align}
Due to \eq{gbig} and \eq{zz1} dependence of $w$ on $E_0$ is  weak unless $\rho_0\gg 1$.

%%%%%
\section{$\jpsi$ dissociation rate }

One of the most interesting applications of the formalism described in the previous sections is calculation of  the dissociation rate of $\jpsi$ which is considered a litmus test of the quark-gluon plasma \cite{Matsui:1986dk}. 

Let $z$ be the heavy ions collision axis; heavy-ion collision geometry implies  that $\bm b_0\cdot \unit z=0$. The plane containing $z$-axis and perpendicular to the magnetic field direction is the reaction plane.  We have
\begin{align}
(\bm b_0\times \bm V)^2&= V_z^2+V_\bot^2\sin^2\phi
\end{align}
where   $\phi$ is the angle between the directions of $\bm B_0$ and $\bm V_\bot$ and we denoted vector components in the $xy$-plane by the subscript $\bot$.
We can express the components of the quarkonium velocity $\bm V$ in terms of the rapidity $\eta$ as $V_z= \tanh \eta$, $V_\bot= P_\bot/(M_\bot\cosh\eta)$, where $\bm P$ and $M$ are the quarkonium momentum and mass and $M_\bot^2= M^2+P_\bot^2$.

%%%%
\begin{figure}[ht]
\begin{tabular}{cc}
      \includegraphics[height=5cm]{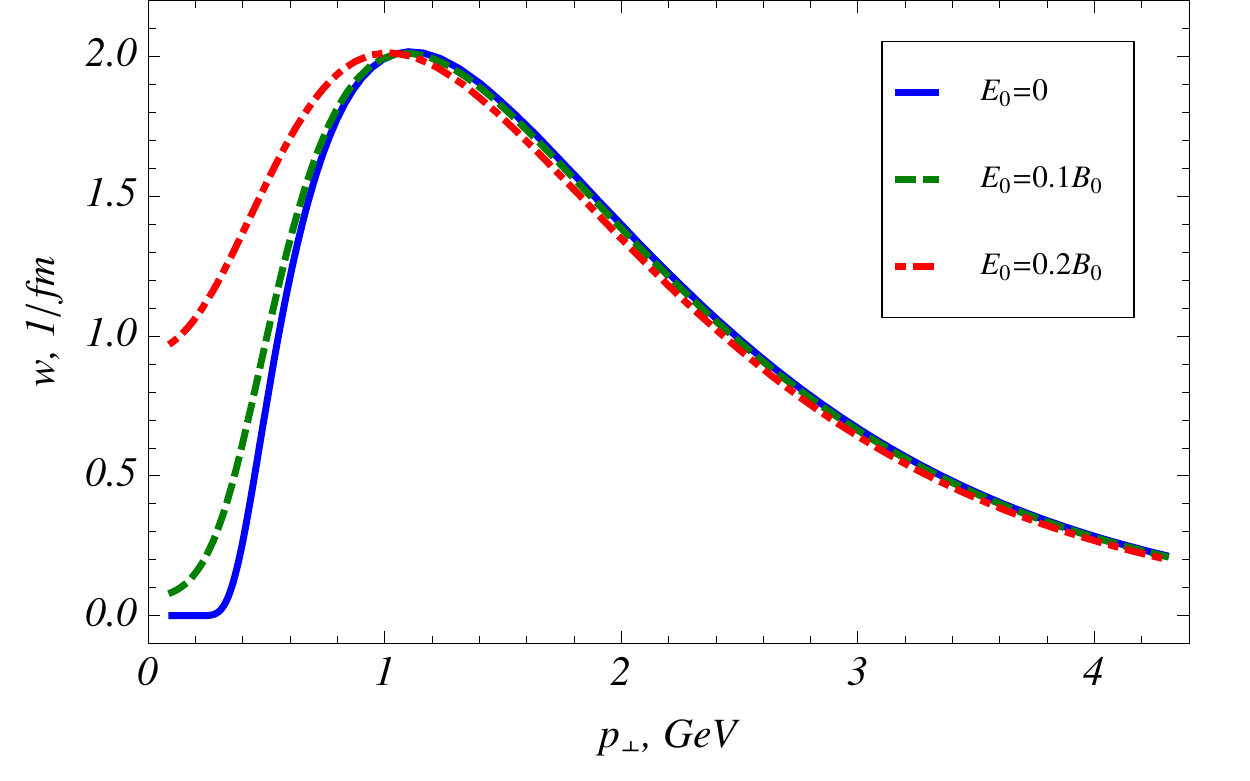} &
      \includegraphics[height=5cm]{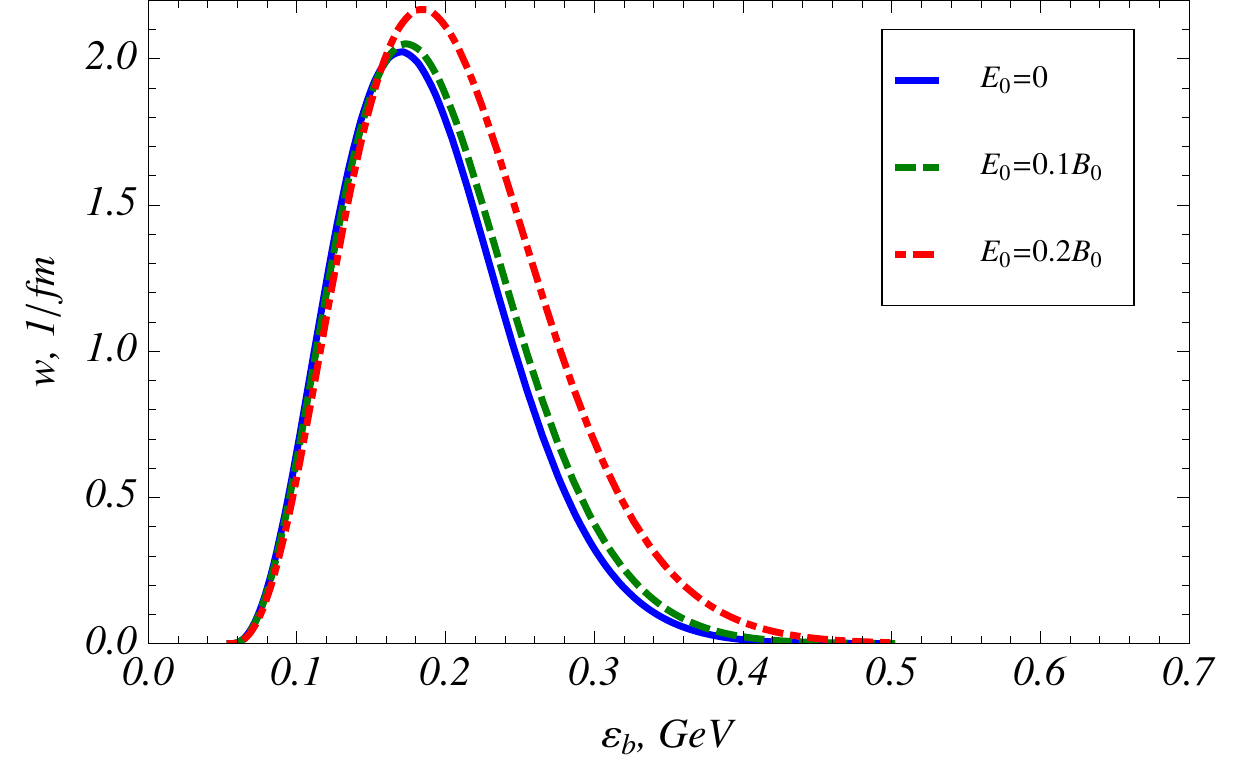}\\
      (a) & (b) 
      \end{tabular}
  \caption{ Dissociation rate of $\jpsi$ at $eB_0=15m_\pi^2$, $\phi=\pi/2$ (in the reaction plane), $\eta=0$ (midrapidity) as a function of (a) $P_\bot$  at $\e_b= 0.16$~GeV and (b) $\e_b$ at $P_\bot=1$~GeV. }
\label{fig:pt}
\end{figure}
\begin{figure}[ht]
      \includegraphics[height=7cm]{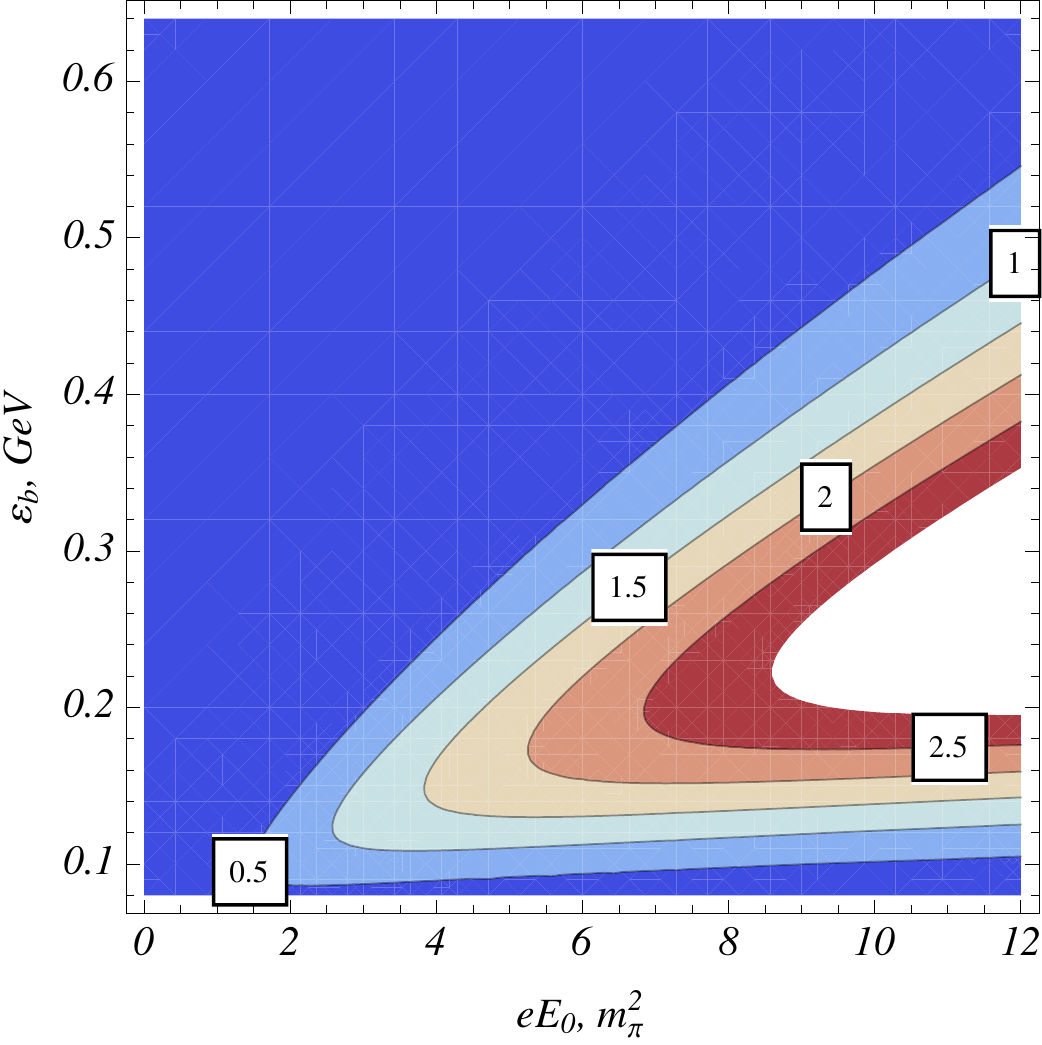} 
  \caption{Contour plot of the dissociation rate of $\jpsi$ as a function of $\e_b$ and $eE_0$ at   $eB_0=15m_\pi^2$, $\phi=\pi/2$ (in the reaction plane), $\eta=0$ (midrapidity) and $P_\bot=0.1$~GeV. Numbers inside boxes indicate the values of $w$ in 1/fm.}
\label{fig:contour}
\end{figure}
\begin{figure}[ht]
\begin{tabular}{cc}
      \includegraphics[height=4.5cm]{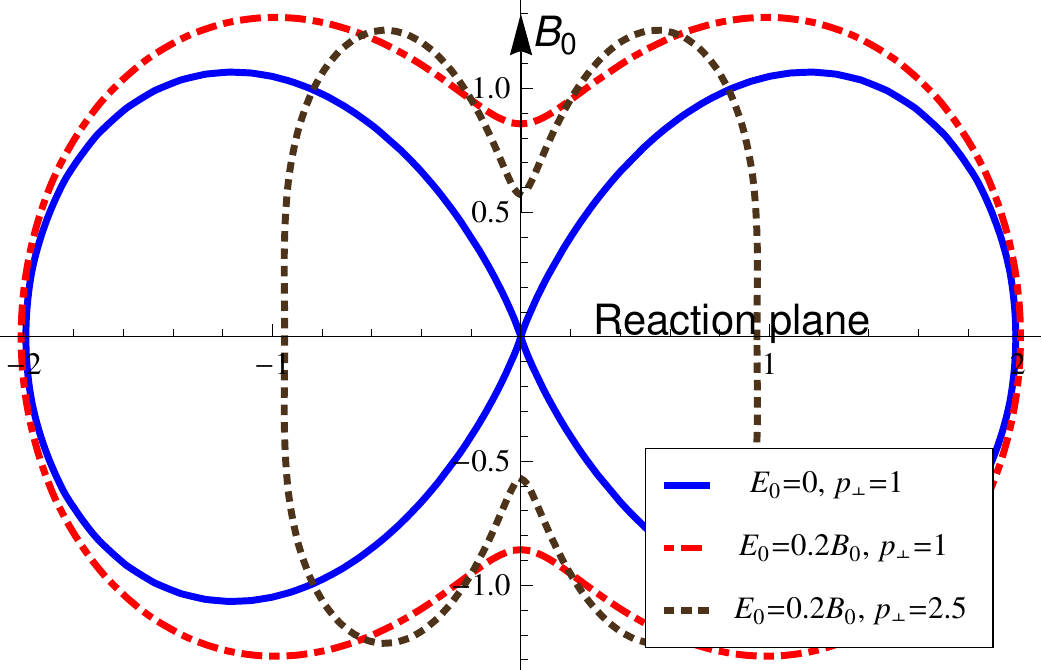} &
      \includegraphics[height=5.cm]{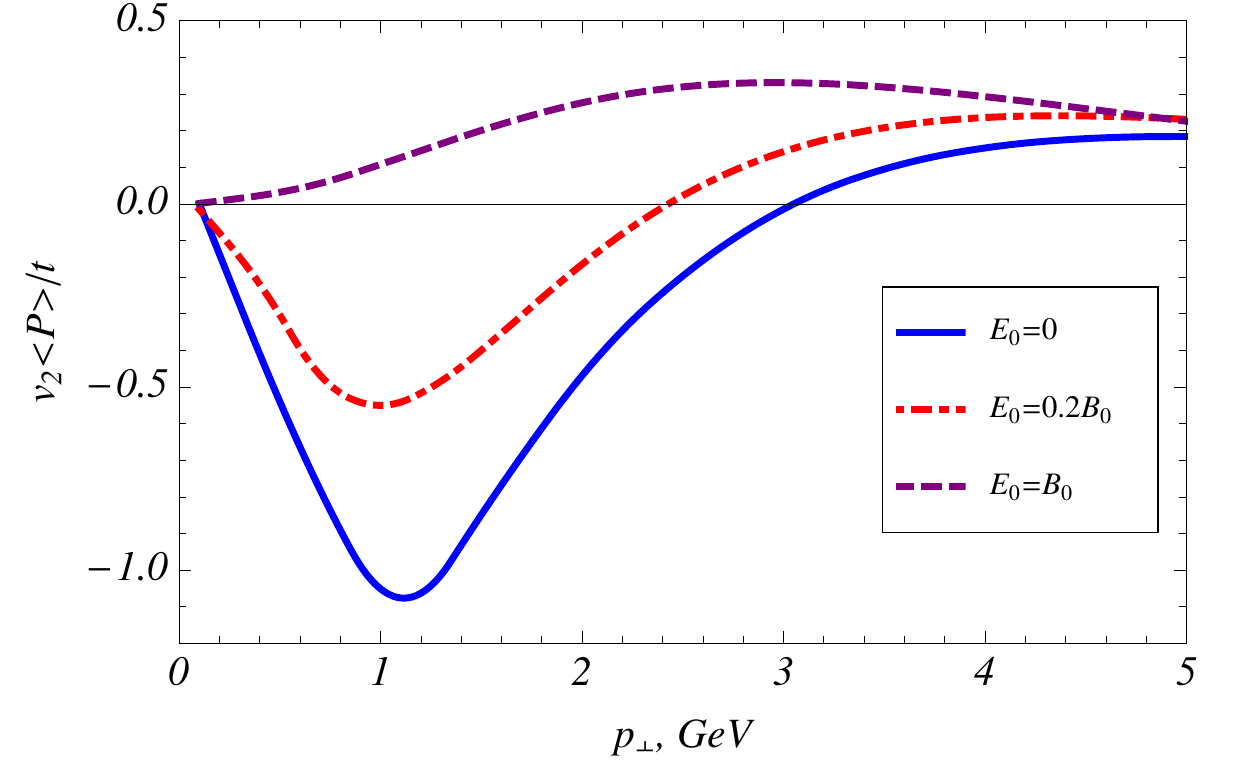}
      \end{tabular}
  \caption{ (a) Angular distribution of $\jpsi$ dissociation rate  at $eB_0=15m_\pi^2$,  $\eta=0$ at different $E_0$ and $P_\bot$ (in GeV's). Magnetic field $\bm B_0$ points in the positive vertical direction. Reaction plane coincides with the horizontal plane. (b) Rescaled second Fourier-harmonic $v_2$  of the azimuthal distribution  as a function of $P_\bot$. $\aver{P}$ is the azimuthal average of the survival probability and $t$ is the time spent by $\jpsi$ in the $P$-odd bubble. }
\label{fig:azimuth}
\end{figure}

%%%%
Results of  numerical calculations are exhibited in Figs.~\ref{fig:pt}--\ref{fig:azimuth}.
In \fig{fig:pt} we show the  dissociation rate of $\jpsi$ for several values of the electric field $\bm E_0$ induced by the Chiral Magnetic Effect.  Note, that the typical size of the medium traversed by a quarkonium is a few fm. Therefore, $w\sim 0.3-0.5$ fm$^{-1}$ corresponds to complete destruction of $\jpsi$'s. 
This means that in the magnetic field of strength $eB_0\sim 15m_\pi^2$ all $\jpsi$'s with $P_\bot\gtrsim0.5$~GeV  are destroyed   independently of the strength of $E_0$. This $P_\bot$ is lower  than we estimated previously in \cite{Marasinghe:2011bt}  neglecting the pre-factors in the dissociation rate. Since  magnetic field strength decreases towards the QGP periphery, most of $\jpsi$ surviving at later times originate from that region. 
Effect of the electric field $\bm E_0$ of the parity-odd bubble  is strongest at low $P_\bot$, which is consistent with our discussion in the previous section. The dissociation rate at low $P_\bot$ exponentially decreases with decrease of $E_0$. Probability of quarkonium ionization by the fields below $E_0\lesssim 0.1 B_0$ (i.e.\ $\rho_0\lesssim 0.1$) is exponentially small. This  is an order of magnitude higher than the estimate $\rho_0\sim \alpha$ proposed in \cite{Kharzeev:2007tn}.

%%%%%%
%%%%%
As the plasma temperature varies, so is the binding energy of quarkonium although the precise form of the function $\e_b(T)$ is model-dependent. 
The dissociation rate picks at some $\e_b^0<\e_b^\text{vac}$ (see \fig{fig:pt}(b)), where $\e_b^\text{vac}$ is the binding energy in vacuum,  indicating that $\jpsi$ breaks down even before $\e_b$ drops to zero, which is the case at $\bm B_0=0$. This $\e_b^0$ is a strong function of $E_0$ as can be seen in \fig{fig:contour}. 
It satisfies the equation $\partial w/\partial \e_b = 0$. In the case $\gamma\ll 1$  \eq{zz} and \eq{zz1} imply 
that 
\beql{e0-1}
\e_b^0= \frac{m}{2}\left( \frac{5eE}{2m^2}\right)^{2/3}\,,\quad \gamma\ll 1
\eeq
At $\gamma\gg 1$ and $\theta= \pi/2$ we employ \eq{gbig} to derive the condition $(\e_b^0)^2+eB\e_b^0/2m-eE^2/B=0$. In view of \eq{gg}  $E\approx B$ and we obtain
\beql{e0-2}
\e_b^0=\frac{eB}{4m}\left( \sqrt{\frac{16m^2}{eB}+1}-1\right)\approx \sqrt{eB}\,,\quad \gamma\gg 1
\eeq
where in the last step we used that $eB\ll m^2$. For a given function $\e_b(T)$ one can convert $\e_b^0$ into the dissociation temperature, which is an important phenomenological parameter.  

%%%%
%%%%%
In the absence of the CME the dissociation probability peaks in the direction perpendicular to the direction of magnetic field $\bm b_0$, i.e.\ in the reaction plane. Dissociation rate vanishes in the $\bm b_0$ direction. Indeed,  for $\bm V\cdot \bm b_0=0$ \eq{em-cf} implies that $E=0$.  This feature is seen in the left panel of \fig{fig:azimuth}. At finite $\bm E_0$ the  dissociation probability is finite in the $\bm b_0$ direction making the azimuthal distribution more symmetric. The shape of the azimuthal distribution strongly depends on quarkonium velocity: while at low $V$ the strongest dissociation is in the direction of the reaction plane, at higher $V$ the maximum shifts towards small angles around the $\bm b_0$ direction. 
Extrema of the azimuthal distribution are roots of the equation $\partial w/\partial \phi=0$. At  $\gamma\ll 1$ it yields minimum at  $\phi_0=0$, maximum at $\phi_0=\pi/2$ and another maximum that satisfies 
the condition (neglecting the spin-dependence of $\e_b$) 
\beql{max=phi}
eE_0\sqrt{1+\gamma_L^2(V_z^2+V_\bot^2\sin^2\phi_0)(1+\rho_0^{-2})}= \frac{2m^2}{3}\left( \frac{2\e_b}{m}\right)^{3/2}
\eeq
In order to satisfy  \eq{max=phi} $\phi_0$ must decrease when $V$ increases and visa versa. This features are seen in the left panel of \fig{fig:azimuth}.

Spectrum of quarkonia  surviving in the electromagnetic field is proportional to the survival probability $P=1-w t$, where $t$ is the time spent by the quarkonium in the field. Consider $P$ as a function of the angle $\chi$ between the quarkonium velocity and the reaction plane $\chi= \pi/2-\phi$. Fourier expansion of $P$ in $\chi$ reads
\beql{four}
P(\chi)= \frac{1}{2}P_0+\sum_{n=1}^\infty P_n\,\cos(n\chi)\,,\quad P_n= \frac{1}{\pi}\int_{-\pi}^{\pi}P(\chi)\,\cos(n\chi)\,d\chi\,.
\eeq
Ellipticity of the distribution is characterized  by the ``elliptic flow" coefficient $v_2$ defined as 
\beq\label{v2}
v_2=\frac{P_2}{\frac{1}{2}P_0}= \frac{\int_{-\pi}^{\pi} (1-wt)\,\cos2\chi\, d\chi}{\pi\aver{P}} = - \frac{t}{\pi\aver{P}}\int_{-\pi}^{\pi}  w\,\cos2\chi \, d\chi
\eeq
where $\aver{P}$ denotes average of $P$ over the azimuthal angle. These formulas are applicable only as long as $wt<1$ because otherwise there are no surviving quarkonia. 
In  the right panel of \fig{fig:azimuth} we show $v_2\aver{P}/t$, which is independent of $t$, as a function of $P_\bot$. As expected, in the absence of the CME, $v_2$ is negative at low $P_\bot$ and positive at high $P_\bot$. $v_2$ changes sign at $P_\bot$ that depends on the strength of the electric field. It decreases as  $E_0$ increases until at $E_0\simeq B_0$ it becomes positive at all $P_\bot$.  
\fig{fig:azimuth}(b) provides the low bound for $v_2$  because $\aver{P}<1$ and $t\gtrsim 1$~fm  . We thus expect that  magnetic field strongly modifies the azimuthal distribution of the produced $\jpsi$'s. Role of the magnetic field in generation of azimuthal anisotropies in heavy-ion collisions has been pointed out before in \cite{Tuchin:2010gx,Mohapatra:2011ku}.

%%%%%%%%
\section{Summary}

In this paper we studied the effect of the parallel electric and magnetic fields on the dissociation rate of quarkonia, and particularly of $\jpsi$. Our main observation is that the CME effect on the dissociation rate is significantly different than the effect of the pure magnetic field if $E_0\gtrsim 0.1 B_0$ that implies  an estimate of the $\Theta$-parameter: $|\Theta|/\pi \simeq  0.1/\alpha$. This is about an order of magnitude larger than is required for the charge separation \cite{Kharzeev:2007tn}. 
Due to the electric field $E_0$ of the parity-odd bubble quarkonium dissociation rate is finite at low $P_\bot$, as indicated in \fig{fig:pt}. (Fortunately, $\jpsi$'s can be measured down to very low $P_\bot$ \cite{Adare:2006ns,Pillot:2011zg}). The effect of the electric field is most clear along the direction perpendicular to the reaction plane, because magnetic component of the Lorentz force vanishes in this direction, see \fig{fig:azimuth}(a). 

Azimuthal distribution of dissociation rate is strongly  asymmetric in external magnetic field. The second harmonic  $v_2$ of the azimuthal distribution of survival probability   is large and negative at  low $P_\bot$, while at high $P_\bot$ it is positive; zero of $v_2$  depends on the relation between the electric and magnetic fields. According to the preliminary experimental data $\jpsi$'s $v_2$ is either small, about a few per cent, or zero \cite{Atomssa:2009ek,Tang:2011kr}.   Absence of such asymmetry in the experimental data may have two reasons. (i) Magnetic field is significantly weaker  and short-lived than suggested in \cite{Kharzeev:2007jp,Tuchin:2010vs}, which is however at odds with the charge separation observations \cite{:2009uh,:2009txa,Ajitanand:2010rc}.   (ii)  Almost none of $\jpsi$'s produced in the center of QGP survive. Rather they originate from the peripheral regions. The later scenario is realized if  time $t$ spent by quarkonium in the field is large  because the dissociation rate only linearly increases with  $t$  but exponentially decreases with the decrease of the field strength toward the QGP periphery. Finally, if the bubble interference effects due to finite bubble size discussed at the end of  \sec{secII} are important they can significantly reduce the $CP$-odd effect  on $\jpsi$ dissociation rate.

Abundance of possible effects associated with strong magnetic field calls for a detailed experimental investigation.

%%%%%%%%%%%%%%%%%%%%%%%%%%%%%%%%
\acknowledgments

This work  was supported in part by the U.S. Department of Energy under Grant No.\ DE-FG02-87ER40371.

%%%%%%%%%%%%%%%%%%%%%%%%%%%%%%%%%%%%%

%%%%%%  APPENDIX %%%%%
\appendix*
\section{}\label{appA}

Here we list functions $C$ and $P$ that appear in \eq{w}.
\begin{widetext} 
\begin{align}
P& = \frac{\gamma^2}{\tau_0}\left[ (\tau_0\coth\tau_0+\frac{\sinh\tau_0\cosh\tau_0}{\tau_0}-2)\sin^2\theta+\sinh^2\tau_0\cos^2\theta\right]^{-1/2}\\
C&= \exp\left[ \ln \frac{\tau_0}{2\gamma}+\int_0^{\tau_0}d\tau \left( \frac{\gamma}{\xi(\tau)}-\frac{1}{\tau_0-\tau}\right)\right]\\
\xi(\tau)&= \left\{ \frac{1}{4}(\tau_0^2-\tau^2)^2\cos^2\theta+\tau_0^2\left[ \left( \frac{\cosh\tau_0-\cos\tau}{\sinh\tau_0}\right)^2-\left( \frac{\sinh\tau}{\sinh\tau_0}-\frac{\tau}{\tau_0}\right)^2\right]\sin^2\theta\right\}^{1/2}
\end{align}
\end{widetext}

\end{document}